\documentclass[conference]{IEEEtran}
\usepackage{cite}

\ifCLASSINFOpdf
\else
  \usepackage[dvips]{graphicx}
\fi
\usepackage[cmex10]{amsmath}
\usepackage{array}
\usepackage[tight,footnotesize]{subfigure}
\usepackage{url}
\usepackage{amsthm}
\usepackage{amssymb}
\theoremstyle{plain}

\theoremstyle{remark}

\begin{document}
\title{Coded Splitting Tree Protocols}

\author{
\IEEEauthorblockN{
Jesper H. S\o rensen, \v Cedomir Stefanovi\' c, and Petar Popovski
}
\IEEEauthorblockA{
Aalborg University, Department of Electronic Systems,
Fredrik Bajers Vej 7, 9220 Aalborg, Denmark
\\
E-mail: \{jhs,petarp,cs\}@es.aau.dk
}
}
\maketitle

\begin{abstract}
This paper presents a novel approach to multiple access control called \textit{coded splitting tree protocol}.
The approach builds on the known tree splitting protocols, code structure and successive interference cancellation (SIC).
Several instances of the tree splitting protocol are initiated, each instance is terminated prematurely and subsequently iterated. The combined set of leaves from all the tree instances can then be viewed as a graph code, which is decodable using belief propagation. The main design problem is determining the order of splitting, which enables successful decoding as early as possible. Evaluations show that the proposed protocol provides considerable gains over the standard tree splitting protocol applying SIC. The improvement comes at the expense of an increased feedback and receiver complexity.  
\end{abstract}
\IEEEpeerreviewmaketitle

\section{Introduction} \label{sec:introduction}

Distributed random access control schemes, like ALOHA\cite{R1975} and splitting-tree protocols\cite{capetanakis}, represent a simple but popular choice for the multiple access channels. In their basic variants, these protocols treat packet collisions as waste, but the recent research has shown that successive interference cancellation (SIC) can significantly increase the throughput of random access protocols. The use of SIC was explored in the splitting-tree framework\cite{SICTA}, doubling the throughput of the binary version of the scheme. Recently, SIC has been used as a main ingredient of the coded random access protocols \cite{casini}:  the users are allowed to transmit packet replicas in multiple slots of the frame and once a transmission has been resolved, SIC is used to remove its replicas, potentially ``unlocking'' the collisions where these replicas may have occurred. Analogies between SIC and iterative belief-propagation (BP) erasure-decoding were established in \cite{liva}, leading to the application of the codes-on-graphs theory to the design of framed ALOHA-based random access schemes. In \cite{SPV2012}, these ideas were extended by applying the concepts of rateless coding to the slotted ALOHA.

In this paper we show how the ideas of coded random access can applied using splitting tree protocols as basis, rather than ALOHA. We propose the \emph{coded splitting tree protocol}, consisting of a set of partially split trees whose combined leaves constitute a graph code, over which the receiver applies SIC. 
The focus of our work is on the \emph{split strategy} that optimizes the distribution of user collisions over the leaves, such that a suitably chosen reward function related to the evolution of the SIC is maximized.
The proposed scheme is derived for the general case when the number of contending users is not a priori known.

The organization of the rest of the text is as follows.
Section~\ref{sec:sysmodel} introduces the system model.
Section~\ref{scheme} presents the main concepts of the coded splitting tree protocol, followed by an analysis in Section~\ref{sec:analysis}. Numerical results are presented in Section~\ref{sec:results}. Finally, Section~\ref{sec:conclusions} concludes the paper.
\section{System Model}
\label{sec:sysmodel}

We consider a multiple access channel with a single receiver and $N$ transmitters, also referred to as users, with the population size $N$ being unknown to all of them.
Link time is organized in fixed-size slots; the duration of the slot is equal to the duration of user messages.
Every time $i-$th user transmits, $1 \leq i \leq N$, he sends a a replica of its message $X_i$
(we assume that each replica contains pointers to all other replicas, as this is required for the execution of SIC).
The transmissions from different uses are assumed synchronized and the received signal in slot $j$, $Y_j=\sum_{i\in A_j}X_i$ is noiseless, where $A_j$ is the set of active (transmitting) users in slot $j$ and their number is denoted by $|A_j|$.
Slot $j$ is referred to as: \textit{idle} if $|A_j|=0$, a \textit{single} if $|A_j|=1$ and a \textit{collision} if $|A_j|>1$. The receiver is only able to distinguish between these three events in a given slot, but is not able to determine the multiplicity of a collision, also referred to as the collision \emph{degree}.
It is assumed that the receiver is able to store any number of $Y_j$ in memory, in order to perform SIC at a later stage.
As an example, if $Y_1=X_2+X_5$ (collision) and $Y_2=X_5$ (single), then $Y_1-Y_2=X_2$, which provides an additional message.
In this way the collisions may potentially be resolved, whereby the corresponding slots contribute to the overall throughput, defined as $T=\frac{R}{M}$, where $R$ is the number of recovered messages and $M$ is the number of used slots.
The system also consists of a feedback channel, through which the receiver can broadcast instructions to the transmitters.
The performance of the system is defined by the achievable throughput under a given number of feedback messages $F$.
\section{Coded Splitting Tree Protocol}\label{scheme}

\begin{figure}[t]
\centering
\includegraphics[width=\columnwidth]{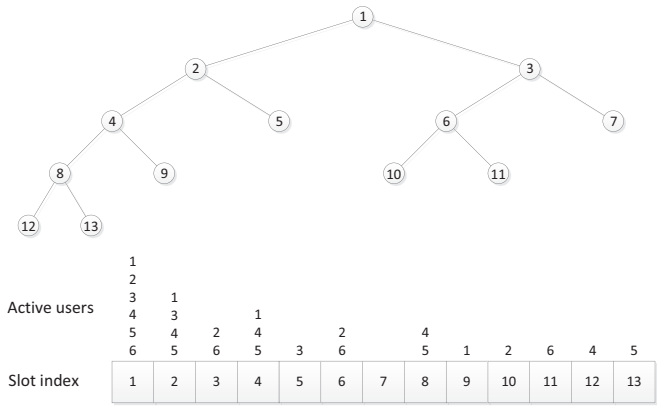}
\caption{Example of the tree splitting algorithm.}
\label{fig:bts_ex}
\end{figure}

In this section, we describe the proposed scheme.
Our approach is centered around the tree splitting algorithm first introduced in \cite{capetanakis}.
In tree splitting, collisions are resolved by asking the involved users to retransmit their message in one of the following $B$ slots with equal probabilities.
This will potentially isolate one or more of the users, whereby their messages are successfully received.
Any remaining collisions are treated in the same way until all collisions are resolved.
This process can be viewed as a tree, where each node represents a received signal, which can either be an idle (0), a single (1) or a collision (c). A collision is the parent of $B$ children, if an attempt to resolve it has been made.
When the process is completed, each leaf in the tree has at most degree 1.
See Fig. \ref{fig:bts_ex} for an example of $N=6$ and $B=2$.
Here the nodes are numbered according the slot number they represent. Under the tree it is seen which users transmit in the individual slots.

This work focuses on the simplest case where $B=2$, which is called binary tree splitting (BTS).
This approach has an asymptotic average throughput of $0.347$ for $N\rightarrow \infty$ \cite{capetanakis}. The use of SIC enables to perform a binary split using only a single slot. For the example on Fig. \ref{fig:bts_ex}, after receiving the signal $Y_2$ in slot $2$, the receiver can locally create the received signal $Y_3$ using $Y_1$ and $Y_2$; thus the slot $3$ can be saved. Using this idea, the average throughput of the splitting tree algorithm has been doubled to $0.693$ \cite{SICTA}.

BTS is a key component of the proposed scheme.
However, instead of completing the algorithm, the tree is only partially split, such that a number of collisions remains.
The resulting leaves will follow a certain degree distribution denoted $\Omega$, where $\Omega(d)$ is the fraction of leaves for which $|A|=d$.
Note that if BTS was fully completed, such that $R=N$, then $\Omega(0)=\frac{M-N}{M}$, $\Omega(1)=\frac{N}{M}$ and $\Omega(d)=0$ for $d>1$, as there are no unresolved collisions.
For partial BTS, $\Omega(d)$ is a random variable for all $d \le N$, where $\Omega(0)$ and $\Omega(1)$ can be observed in the tree.
The choice of splitting strategy, i.e. which collisions to resolve, determines the statistics of $\Omega(d)$.
This relationship is subject to analysis in section \ref{sec:analysis}.

The main idea of coded splitting tree protocol, and the main conceptual contribution of this work, is to generate $K$ partially split trees and view the combined set of leaves as a graph code on the $N$ messages from the users.
Each user transmits its message such that its multiple replicas exist among the leaves, which provides an SIC potential similar to in \cite{casini}. The resulting graph code can thus be decoded using SIC (i.e, iterative BP decoding), as applied in erasure codes like LDPC codes and LT codes.
Hence, existing results in this area can serve as guidelines for the desired $\Omega$ and thereby for the splitting strategy.

Since $N$ is unknown, it is not possible to design an a priori splitting strategy a priori to fulfill the requirements to $\Omega$. We therefore propose a two-phase strategy:

\begin{enumerate}
\item \textbf{Estimation phase:} Perform BTS in each tree, using a predefined strategy, until a predefined point, where the observations in the tree serve as basis for an estimation of the degree of the remaining collisions.
\item \textbf{Degree optimization phase:} Based on the partially split trees and the collision degree estimates from the estimation phase, a strategy for further splitting is chosen, which maximizes a reward function on $\Omega$.
\end{enumerate}

\subsection{Estimation Phase}
In the estimation phase, the goal is to partially split the $K$ trees and estimate the degrees of the remaining collisions. Hence, $N$ is not explicitly estimated, rather how the population has been distributed across the transmission slots, which is more informative and useful in our case. Information for this estimation comes from idle and single slots, since only these have finite entropy in the observation (in fact zero entropy).
The strategy for each tree is to perform the BTS algorithm, with the modification that only a single child is split, if the split of the parent resulted in two collisions.
Once an idle or a single slot is observed, the algorithm is repeated, starting at a collision node closest to the root.
Note that this requires a feedback message, since the users are not aware of the outcome of a transmission slot.
Using the token-based representation for tree protocols from \cite{PFP4004}, it can be seen that any node in the $i$'th level of the tree covers $\frac{1}{2^{i-1}}$ of the probability mass, since each user has an a priori probability of $\frac{1}{2^{i-1}}$ to transmit in the corresponding slot. The described process is iterated until the total fraction of the probability mass covered by idle and single slots exceeds a threshold $\alpha$.
Based on the observations, the degree distributions for the remaining collisions are calculated, as detailed in section \ref{sec:analysis}.

\subsection{Degree Optimization Phase}
The purpose of the degree optimization phase is to perform binary splits in the trees in an order that favors SIC. Hence, we are looking for the sequence of splits, referred to as the \textit{split order}, which modifies the trees resulting from the estimation phase, such that the degrees among the leaves follow a desirable distribution.
The first split, after the estimation phase, is determined by first calculating the resulting degree distribution, $\Omega^1$, for each possible node to be split, $s_1=\{1,2,...,C\}$, where $C=M-\Omega(0)-\Omega(1)$, is the number of collisions among the leaves after the estimation phase.
See details in section \ref{sec:analysis}.
The split that maximizes the scalar product of $\Omega^1$ and the reward function, $\lambda$, is then chosen. The reward function assigns a value to a leaf of any degree $d$, i.e. $\lambda(d)$ is the value of a leaf with degree $d$.
The next splits are chosen in the same manner, using $\Omega^{i-1}$ as the starting point for choosing the split $s_i$.
Note that each split causes $M$ to increase by one.
Once a split order of the desired length has been determined, it is broadcast to the users, which start to transmit according to the schedule.
After each split, belief propagation decoding is attempted, thus giving the whole protocol a flavor of \emph{rateless codes}, as in \cite{SPV2012}.
When all collisions have been resolved, either through a split or SIC, a terminating feedback message is broadcast.

\begin{figure}[t]
\centering
\includegraphics[width=\columnwidth]{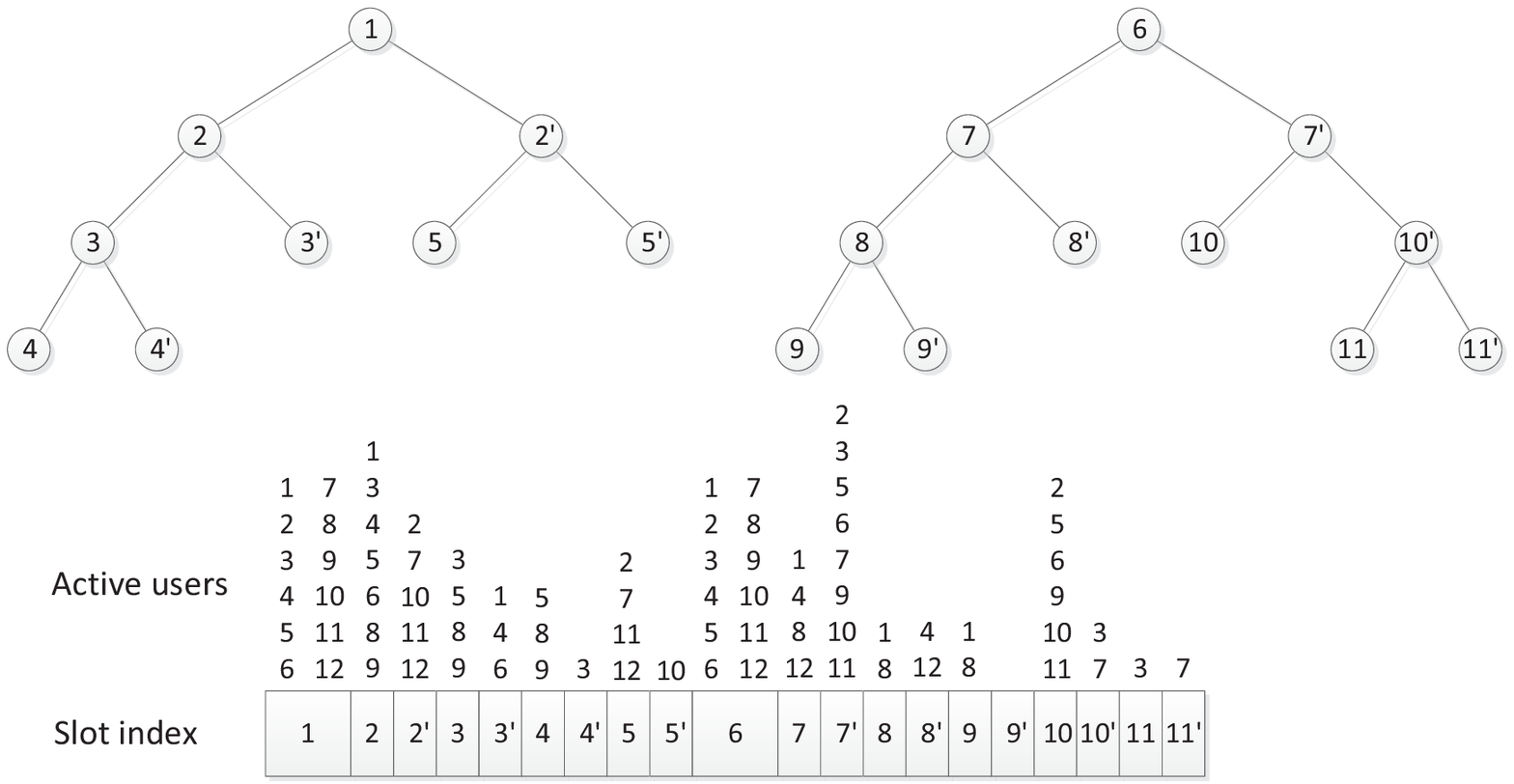}
\caption{Example of the estimation phase.}
\label{fig:matc_ex}
\end{figure}

\subsection{Example}

\begin{figure}[t]
\centering
\includegraphics[width=\columnwidth]{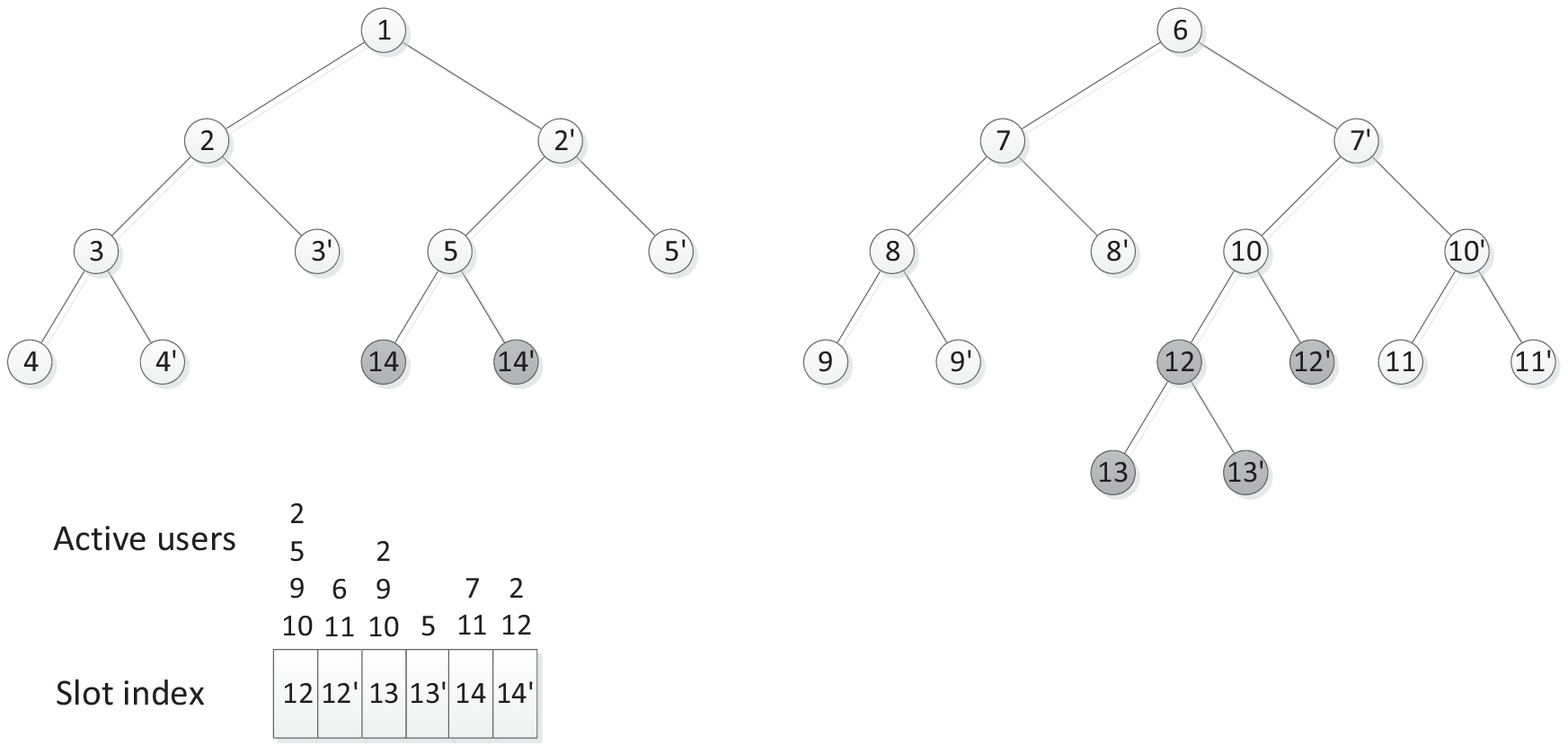}
\caption{Example of the degree optimization phase.}
\label{fig:matc_opt_ex}
\end{figure}

Fig. \ref{fig:matc_ex} and Fig. \ref{fig:matc_opt_ex} show an example of the proposed scheme with $K=2$, $N=12$ and $\alpha=0.2$. The tree nodes are labeled with the slot they represent, which makes it possible to follow the evolution of the trees. Underneath the trees, it is illustrated which users are active in the individual slots and thereby also the degrees of the nodes. Note that since SIC is utilized, only a single slot is necessary for each split, such that the complementary node of the node representing the $i$-th slot is denoted by $i'$.
In other words, node $i'$ can be obtained from $i$ and the parent node, and thus a slot corresponding to node $i'$ is not needed.

In Fig. \ref{fig:matc_ex} the estimation phase is illustrated. In the left tree the first single or idle slot is observed in slot $4'$, after three splits from the root, which means the slot covers a fraction of $\frac{1}{2^3}=0.125$ of the probability mass. Since $\alpha=0.2$ is applied, the estimation phase continues, starting from the node closest to the root, $2'$. After only a single split, another single or idle slot is observed, this time in the third level of the tree. Hence, the node covers $\frac{1}{2^2}=0.25$ of the probability mass, bringing the total mass covered by single and idle slots to $0.375$, which exceeds $\alpha$ and the estimation phase of this tree is terminated. The same procedure is performed in the right tree, where single or idle slots are observed in $9'$, $11$ and $11'$, who cover a total of $\frac{1}{2^3}+\frac{1}{2^3}+\frac{1}{2^3}=0.375$ of the probability mass. Based on the observations in the two trees, the degree distributions of the remaining collisions are calculated as described in section \ref{sec:analysis}.

The degree optimization phase is illustrated in Fig. \ref{fig:matc_opt_ex}. This phase continues on the trees resulting from the estimation phase. It is assumed for this example that splitting node $10$ will maximize the scalar product of $\Omega^1$ and $\lambda$, which means this node is scheduled to be split first. Following this, a split of node $12$ is assumed to maximize the scalar product, and finally node $5$. Hence, the determined split order is $10$, $12$ and $5$, which is broadcast to the users, who flip coins and transmit accordingly. The splits performed in the degree optimization phase are marked with gray in Fig. \ref{fig:matc_opt_ex}. After each split, SIC is attempted.

\section{Analysis} \label{sec:analysis}
Two elements in the proposed scheme are subject to analysis.
First, the leaf degree probability distribution is derived on the basis of an observation of the tree. This is the essence of the estimation phase. Second, the expressions necessary to optimize the split order for the degree optimization phase are derived.

\subsection{Leaf Degree Probability Distribution}
We introduce $\mathbf{D}=\{D_1,D_2,...,D_M\}$, which is a vector of random variables, where $D_{\ell}$ is the degree of the $\ell$-th leaf, $\ell=1,2,...,M$.
Note that given a population $N$, $\mathbf{D}$ follows a multinomial distribution, denoted $\mathcal{M}$, with parameters $N$ and probability vector $\mathbf{P}=\{P_1,P_2,...,P_M\}$, where $P_{\ell}$ is the fraction of probability mass covered by the $\ell$-th leaf, $\ell=1,2,...,M$.
Moreover, we introduce $\mathbb{D}_{\hat{N}}$, which is the set of possible realizations of $\mathbf{D}$ as constrained by the observations in the tree and an estimate of the population $N=\hat{N}$.
Hence, for all $\mathbf{D} \in \mathbb{D}_{\hat{N}}$, we have that $\sum_{\ell} D_{\ell} = \hat{N}$. Also, $D_{\ell} = |A_{\ell}|$, if $|A_{\ell}| < 2$; otherwise $D_{\ell}$ is unknown (i.e, unobserved). 

For the unobserved leaf degrees, we find the probability distribution, $Pr(D_{\ell}=d)$ for $d \ge 2$, by summing the probabilities of all realizations in $\mathbb{D}_{\hat{N}}$ where $D_{\ell}=d$.
Since we have no prior distribution on $\hat{N}$, we must also sum over $\hat{N}=1,2,...,\infty$, weighting each $\hat{N}$ equally:
\begin{align}\label{leaf_dist}
Pr(D_{\ell}=d) &= \sum_{\hat{N}=1}^{\infty} \sum_{\mathbf{D}\in \mathbb{D}_{\hat{N}}:D_{\ell}=d} \mathcal{M}(
\mathbf{D},\hat{N},\mathbf{P}).
\end{align}

The sum in \eqref{leaf_dist} converges whenever at least one idle or single slot is observed, since in that case $\sum_{\mathbf{D}\in \mathbb{D}_{\hat{N}}:D_{\ell}=d} \mathcal{M}(\mathbf{D},\hat{N},\mathbf{P}) \rightarrow 0$ for $\hat{N} \rightarrow \infty$. For that reason, it is in practice enough to sum until a sufficiently high $\hat{N}$, such that convergence is achieved.
The overall leaf degree probability distribution is now found as:
\begin{align}\label{leafdeg}
\Omega(d)=\sum_{\ell=1}^{M} Pr(D_{\ell}=d)\text{ for } d > 0.
\end{align}

\subsection{Split Order Selection}

The degree probability distribution of the individual leaves and the resulting $\Omega(d)$ constitute the starting point of the degree optimization phase.
For this phase we are interested in the optimal split order, across all $K$ trees, according to a reward function $\lambda$, for the remainder of the transmission.
We will add a superscript $i$ to the derived distributions, which denotes the number of performed splits in the degree optimization phase.
Moreover, we concatenate the random variables from the individual trees in a single vector, $\mathbf{D}^0$, representing all leaves.
Hence, $\mathbf{D}^0=\{\mathbf{D}_1,...,\mathbf{D}_k,...,\mathbf{D}_K\}$, where $D_{k,\ell}$ is the $\ell$-th leaf of the $k$-th tree.

The result of a split is that one leaf is replaced by two children.
Hence, if e.g. the $\ell$-th leaf is chosen for the first split, then $D^0_{\ell}$ will be replaced by two new random variables, denoted $D^1_{\ell}$ and $D^1_{\ell+1}$.
For the $i$-th split, we are interested in finding the leaf that maximizes the scalar product of the resulting $\Omega^i$ and $\lambda$.
Hence, if $s_i$ denotes the index of the leaf for the $i$-th split, then the following optimization is performed:
\begin{align}
\underset{s_i}{\text{maximize}} \sum_{d=0}^{\infty} \Omega^i(d)\lambda(d),
\end{align}
where the dependency on $s_i$ lies in $\Omega^i(d)$, since the choice of split determines how $\Omega^{i-1}(d)$ maps into $\Omega^i(d)$. As in \eqref{leafdeg}, we find the overall leaf degree distribution by summing the contributions from the individual leaves:
\begin{align}
\Omega^i(d)&=\sum_{\ell=1}^{M} Pr(D_\ell^{i}=d) \text{ for } d > 0.
\end{align}
Leaves not chosen for the split are unaffected. However, the index of the leaves following the chosen leaf in the vector $\mathbf{D}^i$ is increased by one, since a split causes one leaf to be replaced by two new leaves. Hence,
\begin{align}\label{pDd}
Pr(D_\ell^{i}=d)&=Pr(D_\ell^{i-1}=d) \text{ for } \ell < s_i, \notag \\
Pr(D_\ell^{i}=d)&=Pr(D_{\ell-1}^{i-1}=d) \text{ for } \ell > s_i+1.
\end{align}

Conditioned on the degree of the chosen leaf, $D_{s_i}^{i-1}=d_p$, the new leaves will both follow the binomial distribution, denoted $\mathcal{B}$, with parameters $d_p$ and $0.5$. The probability distribution of the degree of the chosen leaf is found using \eqref{leaf_dist}. We thus have:
\begin{align}
Pr(D_{s_i}^i=d) &= Pr(D_{s_i+1}^i=d) \notag \\
                &= \sum_{d_p=d}^{\infty} \left( Pr(D_{s_i}^{i-1}=d_p) \mathcal{B}(d,d_p,0.5) \right),
\end{align}
which concludes the analysis.
\section{Numerical Results} \label{sec:results}
The proposed scheme has been evaluated and compared to the BTS algorithm.
BTS is applied on a single tree only and at one level of the tree at a time.
Hence, initially the root is split, resulting in a maximum of two remaining collisions.
A feedback message reports which leaves are in collision and must be further split.
This process continues until all collisions have been resolved. This constitutes the reference scheme.

For the coded splitting tree protocol, the reward function, $\lambda$, is chosen to be:
\begin{align}
\lambda(d) = \left\{
\begin{array}{c c}
0.5 & \text{ for } d=2,3, \\
0 & \text{ elsewhere}.
\end{array} \right.
\end{align}

The motivation behind this choice is the dominance of degrees two and three in well performing degree distributions for LT codes \cite{lt,sorensen}.
Experiments show that the performance of the proposed scheme is limited by the amount of leaves with these degrees.
Hence, maximizing the amount of leaves with degrees two and three seems reasonable. Note that using such a reward function will not provide trees containing only leaves with degrees two and three, due to the randomness of the scheme. For this reason, leaves with degree one will occur, as is necessary in order to initiate SIC (i.e., iterative BP erasure-decoding).
Further investigation in the choice of reward function is subject to future work.

The length of the split order, applied in the degree optimization phase, is determined by the point, where an increase in the scalar product is no longer possible. From this point, the leaf with the highest expected degree is chosen for the next split, until decoding succeeds.

\begin{figure}[t]
\centering
\includegraphics[width=\columnwidth]{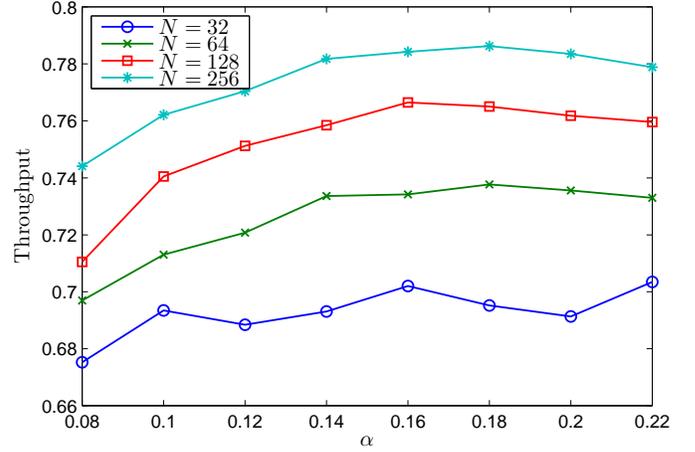}
\caption{Throughput as a function of $\alpha$ for $K=3$.}
\label{fig:K3_th}
\end{figure}

\begin{figure}[t]
\centering
\includegraphics[width=\columnwidth]{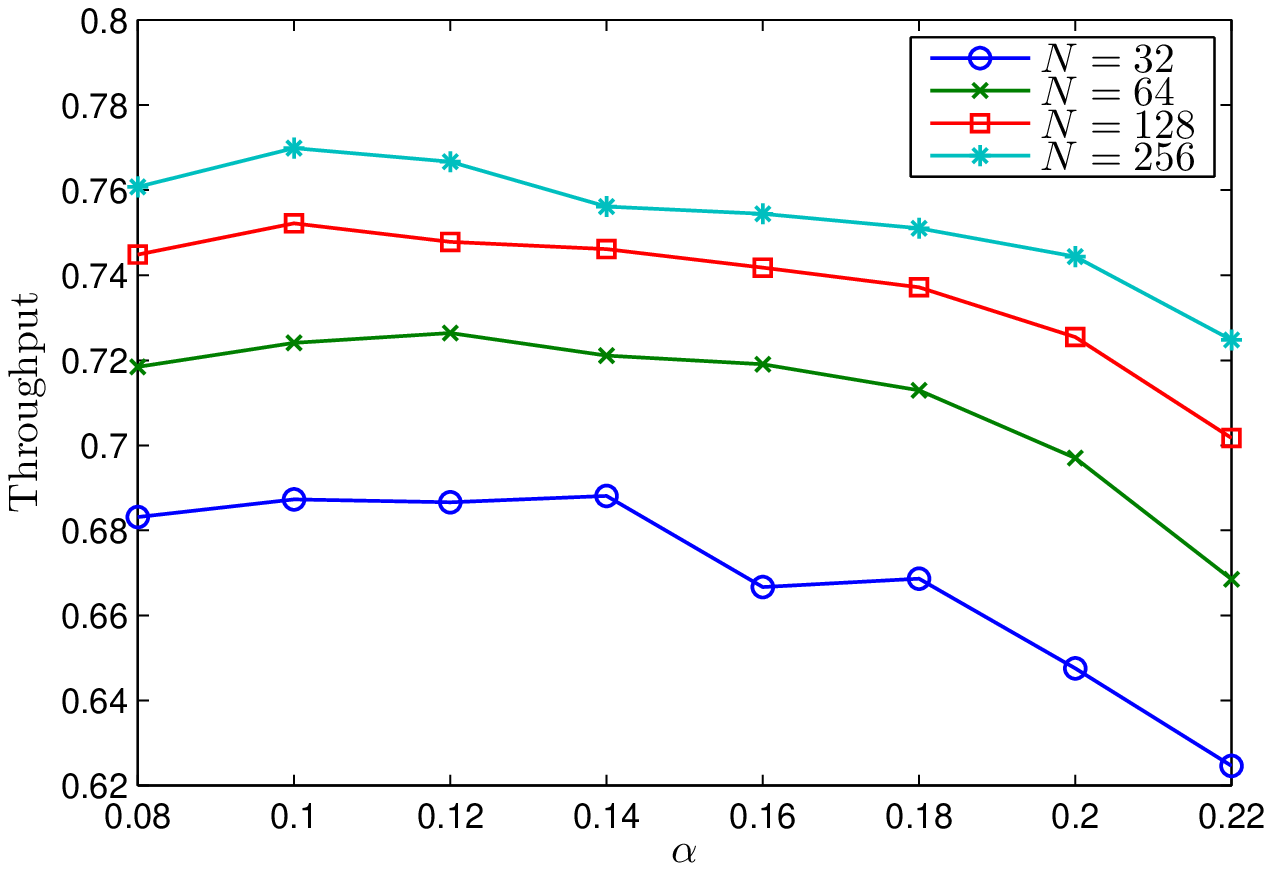}
\caption{Throughput as a function of $\alpha$ for $K=4$.}
\label{fig:K4_th}
\end{figure}

\begin{figure}[t]
\centering
\includegraphics[width=\columnwidth]{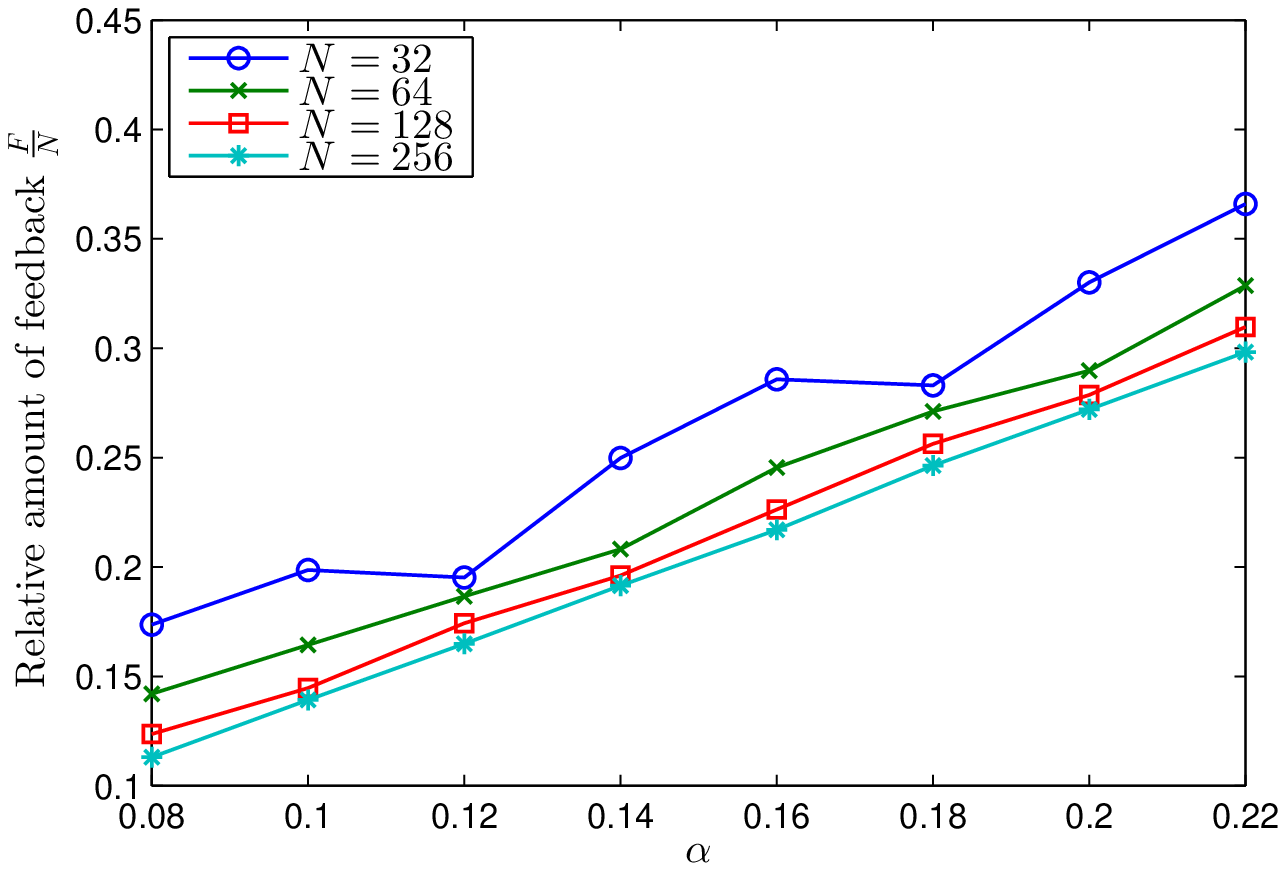}
\caption{The relative amount of feedback as a function of $\alpha$ for $K=3$.}
\label{fig:K3_fb}
\end{figure}

\begin{figure}[t]
\centering
\includegraphics[width=\columnwidth]{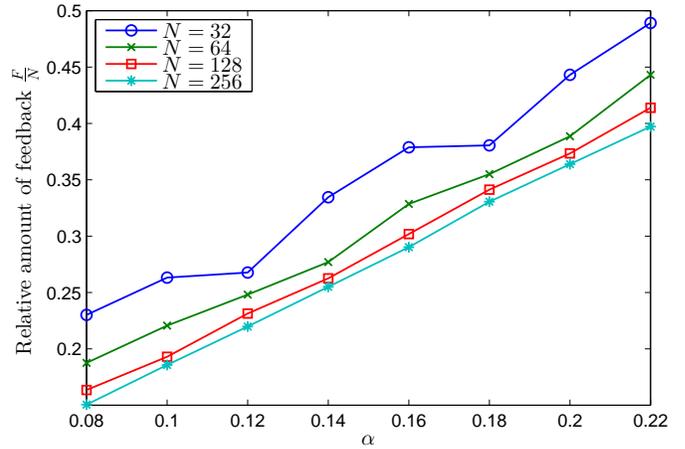}
\caption{The relative amount of feedback as a function of $\alpha$ for $K=4$.}
\label{fig:K4_fb}
\end{figure}

Fig. \ref{fig:K3_th} and Fig. \ref{fig:K4_th} show the achieved throughput as a function of $\alpha$ for $K=\{3,4\}$ and $N=\{32,64,128,256\}$. The number of necessary feedback messages is plotted in Fig. \ref{fig:K3_fb} and Fig. \ref{fig:K4_fb} for the same choices of parameters. It is seen that the optimal choice of $\alpha$ depends on $K$ but not on $N$. Moreover it is seen that $K=3$ provides the better results, which shows that only very few trees are necessary in order to have a SIC potential.

\begin{figure}[t]
\centering
\includegraphics[width=\columnwidth]{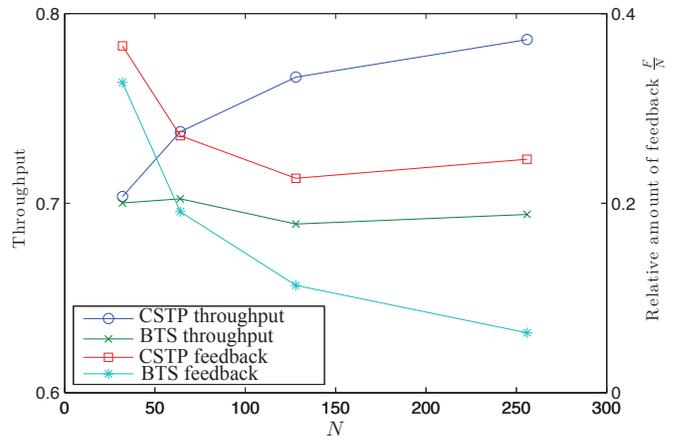}
\caption{Comparison of BTS and coded splitting tree protocol (CSTP) for $K=3$ and optimized $\alpha$.}
\label{fig:comp}
\end{figure}

A comparison with BTS is shown in Fig. \ref{fig:comp} for $K=3$ and $N=\{32,64,128,256\}$, where optimized $\alpha$ has been used for the coded splitting tree scheme, according to the results in Fig. \ref{fig:K3_th}. It is seen that the proposed scheme outperforms BTS with respect to throughput at all $N$ and that the improvement increases with $N$. This is in line with the efficiency increase as a function of the message length seen for erasure codes such as LT codes. Fig. \ref{fig:comp} also shows that the performance improvement comes at the price of increased feedback.

\section{Conclusions}\label{sec:conclusions}
A novel approach to multiple access control called coded splitting tree protocol has been presented, which uses splitting trees and successive interference cancellation in order to create a coded random access. The protocol works by constructing several binary splitting trees, which are terminated prematurely, such that collisions remain among the leaves. The combined set of leaves is then viewed as a graph code, which can be decoded through belief propagation. The key design element is to choose a sequence of splits, which ensures a leaf degree distribution, which favors belief propagation. A design example has been presented, which achieves throughputs close to $0.8$, significantly outperforming the existing tree splitting protocol with SIC. This improvement comes at the price of increased feedback.

\bibliographystyle{ieeetr}
\bibliography{bibliography}

\end{document}